# Next Generation of TCAD Environments for MEMS Design

U. Triltsch, S. Büttgenbach
Technische Universität Braunschweig, Institute for Microtechnology
Alte Salzdahlumer Str. 203
38124 Braunschweig, Germany

*Abstract*-Most available MEMS design environments focus on the integrated behavioral simulation of electronic and mechanical components. The fabrication process is considered a fixed sequence of well-known steps, which can be purchased from foundries. Simulation models rely on the assumption of ideal process conditions and do not take into account tolerances or intra die variations.

Recently there has been discussion that the gap between actual process results and that of predicted models are often out of a tolerable range. This leads to the demand of increased simulation accuracy for the most crucial process steps. The deviations of the comb structures in a resonator, which result from the tolerances in a Deep-RIE process, for example, can make the overall system design fail. To avoid these kinds of errors two things are needed: the technology provider has to know the parameter range of his process and be able to provide it to the designer without loosing his intellectual property. The designer in turn must be provided with a way to use this information to design a system which can perform in the range expected.

In this paper we present the latest version of the TCAD environment BICEP³S (Braunschweigs Integrated CAD-Environment for Product Planning and Process Simulation). By using a central process database, which allows the exchange of all relevant process data it is able to overcome many of the mentioned obstacles. The database and process planning tool can be used by process developers to document changes in process settings and the influence of such changes on the process result. This information can then be used by the designers to set-up a simulation file for a detailed analysis of the impact of such parameter changes on the requested design. This will be shown by the example of silicon etching using an atomistic etch simulator.

I. INTRODUCTION

In this paper the most recent results of the design environment BICEP³S (Braunschweigs Integrated CAD-Environment for Product Planning and Process Simulation) are presented. This TCAD tool consists of a series of modules [1] that integrate, for example, SoftMEMS system-level tools with detailed process simulators such as the etch simulator SUZANA [2] as well as process design tools with a link to process-, material- and media databases [3]. This approach combines system-level behavioral simulation and component-level design tools to provide engineers with a complete view of the overall design process.

Most of today's MEMS-EDA / TCAD environments focus on integrated behavioral simulation of electronic and mechanical components [4, 5]. Fabrication technologies are considered to be a fixed sequence of well known process steps, which cannot be varied. Therefore, the main focus of the designer is on adapting the layout needed to meet the desired performance requirements of a device. There have been many academic publications regarding behavioral modeling and system-level design approaches around the turn of the millennium [6, 7, 8].

Recently there has been discussion that the gap between actual process results and that of predicted models are often out of a tolerable range. This leads to the demand of increased simulation accuracy for the most crucial process steps. The deviations of the comb structures in a resonator, which result from the tolerances in a Deep-RIE process, for example, can make the overall system design fail. To avoid these kinds of errors two things are needed: the technology provider has to know the parameter range of his process and be able to provide it to the designer without loosing his intellectual property. The designer in turn must be provided with a way to use this information to design a system which can perform in the range expected.

II. DESIGN ENVIRONMENT BICEP³S

The design environment BICEP³S focuses on the definition of interfaces between single modules and acts as a central control and data management platform for different phases of microsystem design (figure 1). A central database management system makes sure that all the relevant data, which is obtained during a product design process, is stored in a consistent way and can be accessed by different designers in a simultaneous engineering environment.

Each module focuses on a specialized task in component-oriented design cycles. Special consideration was given to the planning and optimization of consistent process flows for systems, which consist of components that have been developed using different technologies and are integrated into the same innovative microsystem. Once the system is subdivided into functional elements the designer can search the database of predefined components for previously designed parts which meet the functional criteria, as described in [9]. Already existing components can be selected and their stored fabrication data as well as the parameterized layout data is used for the definition of the new component. In other cases components might have to be designed in detail, whereby the layout and process design module will support the designer during this design state.

When a 3D-model without detailed process features is sufficient, the layout and process data can be exported to commercially available 3D modelers, e.g. SoftMEMS or Coventor Designer, where a model can be created for further computational analysis. If a more detailed view of the process result is needed for certain process steps, specialized technology simulator can be used. These process simulators are able to extract parameters from the definition of the process flow and are therefore directly linked to the process design module. If the simulation results are to be used for further computational analysis, there are converters available to ensure conversion between CAD or FEM tools and the simulator. Such is the case for etch simulation results.

In some cases the mask design needs to be optimized. In theses situations the design environment offers a powerful module for the optimization of lithographic masks. An evolutionary algorithm is used to create suitable layouts. These layouts evolve over a certain number of generations starting from a predefined set of masks and a 3D reference structure. The process simulators are used for the assessment of each generated layout and a score is calculated from the predefined weights and degree of matching. To prevent the process from running into a local optimum, mutation is used to generate new individuals [10].

Finally, the process planning module acts as an interface for the collection and documentation of all relevant process results during the fabrication process. For example, process engineers can store measurement protocols during the setup of the process. Such protocols are then used during inline measurements by the operator. This ensures full control over the quality criteria for the executed fabrication steps.

The connection and communication between all single modules is controlled by a workflow system. This system leads the engineers through different steps, which are associated with the current design problem.

As the performance of a system is often dependent on the actual process result, focus will be given to the capabilities of process simulation in the next section of this paper. Wet chemical etching of silicon is still an important technology for MEMS and the built in simulator along with its model

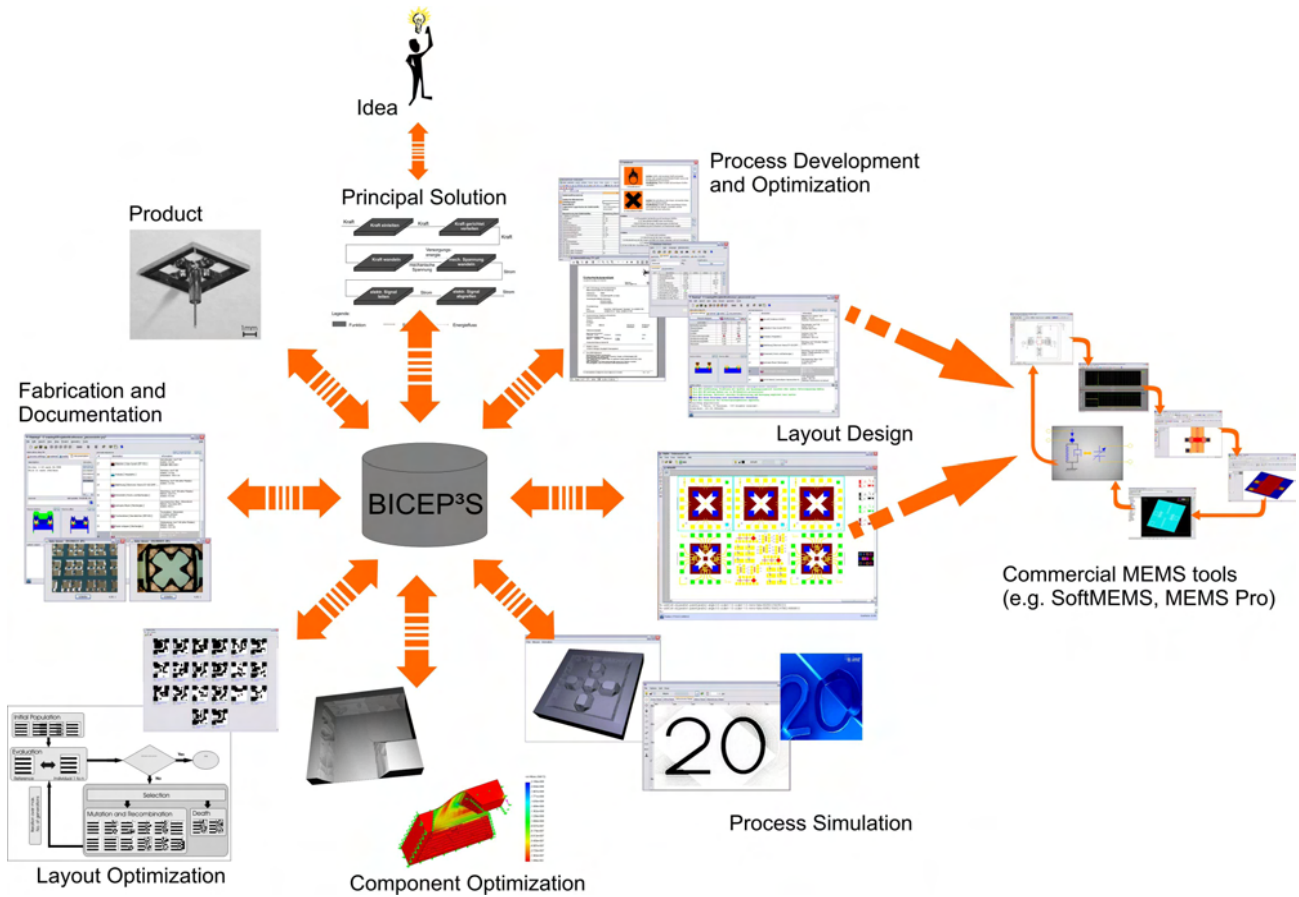

Fig. 1: The single modules of BICEP³S form a closed environment for MEMS design.

## III. ETCH SIMULATION SUZANA

As one example of the process simulation capabilities of BICEP³S, the etch simulator SUZANA is presented. SUZANA uses a cellular automaton model to simulate a wide range of three-dimensional geometries, including etch-stop layers, double side etching and even locally disordered crystal zones. The first version of this simulator was developed from 1991 to 1994 [2]. In the past few years the model has been further enhanced and new features such as tolerance analysis of crucial parameters have been introduced [1]. This has made SUZANA one of the most accurate and flexible tools in this area available today. However, as it considers only the etch rates of the most relevant crystal planes, the simulation might not be as accurate in terms of predicting surface properties on very small features. A more accurate model which takes higher indexed planes into account was presented by Gosálvez [11] and implemented into the VisualTAPAS [12] project. For most design tasks the use of SUZANA is absolutely sufficient and benefits from its fast computation times and accuracy. Figure 2 demonstrates how a comb-like compensation element is removed along the etch limiting (111) planes. The pictures on the right hand side show simulation results after etching with KOH (30%) at 80 °C. The pictures on the left hand side are SEM micrographs. This example shows the high degree of matching between simulation results and the etched structures.

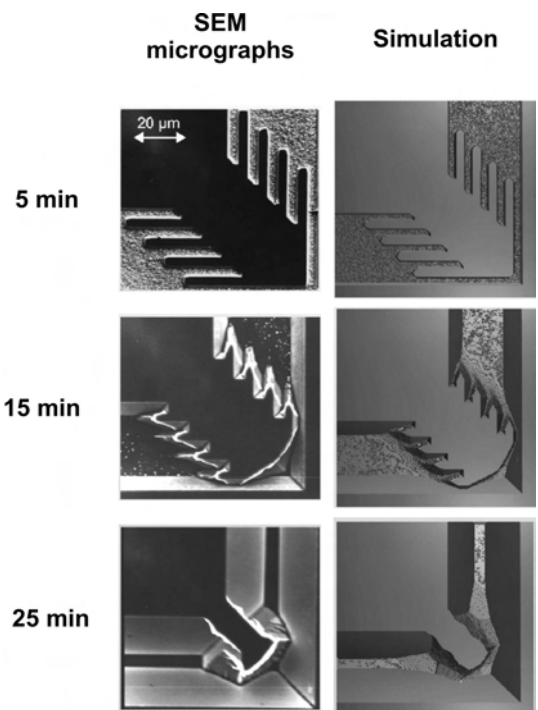

Fig. 2. A comparison between simulation and experimental results shows the high accuracy of the algorithm.

### A. *Simulation Model*

Two different types of etch simulators have emerged during the last few years. One uses the geometric model based on the construction of Wulff-Jaccodine, the other is an atomistically modeled etch process. The delimiting problem of the Wulff-Jaccodine algorithm is the occurrence of saddle points whenever concave-convex shapes are mixed. Frühauf [13] tries to overcome these problems by performing two-dimensional calculations in a first step and calculating three-dimensional sidewalls from the generated line structures. However, simulations for complex undercut geometries and processes involving etching from top and bottom sides are not possible.

SUZANA uses the atomistic approach which implements a cellular automata based model of the etch process. Cellular automata can be considered discrete structures which consist of a large number of simple cells. Such cells are ordered into a grid, interact locally and can represent a certain number of predefined states (in this case etched or not etched). The interaction of all cells is defined by rules. A rule can be regarded as a function, whose arguments are the states of the cell under consideration and of the neighboring cells. Its value is the new state of the considered cell. The application of the rules to a certain cell results in a change of its state.

One basic cell of the silicon crystal structure can be described by two face-centered cubes, which are displaced by a quarter of the silicon lattice constant in each spatial direction. This results in eight relevant atoms for the complete description of one cell. In Figure 3 these atoms are depicted as darker and linked.

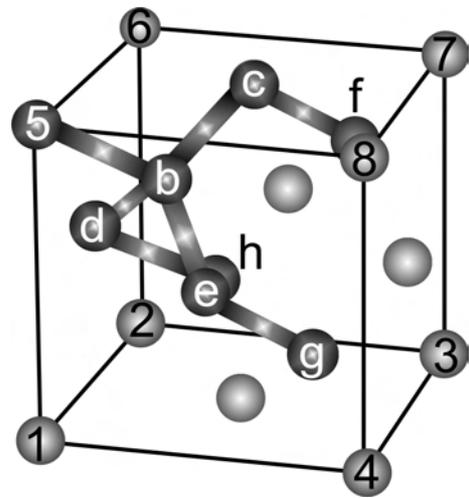

Fig. 3. The elementary cell of the silicon crystal can be modeled by 8 atoms.

There are two models available. If a less detailed model is sufficient, one macro cell of the model is formed by the eight corner points of one cubic cell (1-8). In this case each of the corner atoms represents the eight atoms (5, b-h). If a more detailed model is required a full set of atoms is defined, i.e. one macro cell describes the atoms 5 and b to h. The second model ensures accurate results in multi step simulations and leads to a higher resolution in the z-direction.

As described above, the interaction of the cells is described by rules. The next section will explain how rules model the behavior of the cells in the case of wet-chemical etching of silicon.

*B. Rules for Wet-Chemical Etching*

The simulation uses discrete time steps to calculate the propagation of the etch front. For each step the algorithm iterates over all cells on the surface of the model and calculates the number of neighbors of the current cell. A Depending on the number of neighbors a decision is made as to whether a cell belongs to a certain crystal plane and the etch probability is calculated accordingly. For the three main planes the rules are as follows:

(100)-plane: Atoms of this plane have two neighbors of first order (fig. 5 a) and are etched with a probability $P_{100}$.

(110)-plane: Atoms of this plane have 3 neighbors of first order and less than 9 neighbors of second order OR 9 neighbors of second order and less then 3 neighbors of third order. They are etched with a probability $P_{110}$.

(111)-plane: Atoms of this plane have 3 neighbors of first order and a minimum of 10 neighbors of second order OR nine neighbors of second order and a minimum of 9 neighbors of third order. They are etched with a probability $P_{111}$.

If higher indexed planes have to be taken into account, appropriate rules have to be defined and etch probabilities must be gained from experimental data. Simulation speed decreases with an increasing number of rules.

*C. Parameter and Layout Definition*

The simulation process can be set-up in a graphical user interface. Masks for top and bottom sides as well as etch stop areas can be imported from EDA/ CAD environments or can be created from scratch.

The process panel is used to define dimensions and resolution of the simulated area. Other specific parameters such as etch rates, substrate orientation or concentration and temperature of the etchant can be specified or imported from the central database.

*D. Post Processing of Simulation Results*

To gain full advantage of the simulation results a link to CAD or FEM tools has been developed. The simulation results are stored in discrete voxel file at the specified model resolution. A user defined number of etch states is saved which, if chosen, can be translated into a surface data model and displayed in a built-in 3-D viewer. Another important feature, which speeds-up the design process, is the functionality of ACIS format export. For further use in FEM analysis the detailed simulation results can be simplified. If, for example, the surface topography of sidewalls is of less relevance than the transmission between a membrane and a boss structure as depicted in figure 8, these details can be removed. The ACIS file can be imported into most solid modelers. In the example below, SolidWorks was used to import the simulation result and CosmosWorks was utilized to simulate the mechanical behavior of the simulated structure. Figure 5 gives an overview of the workflow for SUZANA to FEM export.

IV. SUMMARY AND PERSPECTIVE

A TCAD tool for MEMS, which combines process and layout design and provides links to behavioral modeling tools as well as specialized modules for the optimized design of single components was presented. This tool supports the design of new fabrication processes along with behavioral simulation and layout design. The features of the latest version of the modular software environment BICEP³S (Braunschweigs Integrated CAD-Environment for Product Planning Process Simulation) was explained. Due to the importance of detailed simulation of single process steps in fabrication oriented design cycles, an atomistic etch-simulator was presented as an example for specialized simulation modules of our design environment.

One major issue of future research will be the enhancement of the etch simulation software. Currently the implementation of a new cellular automaton for deep reactive ion-etching is being investigated.

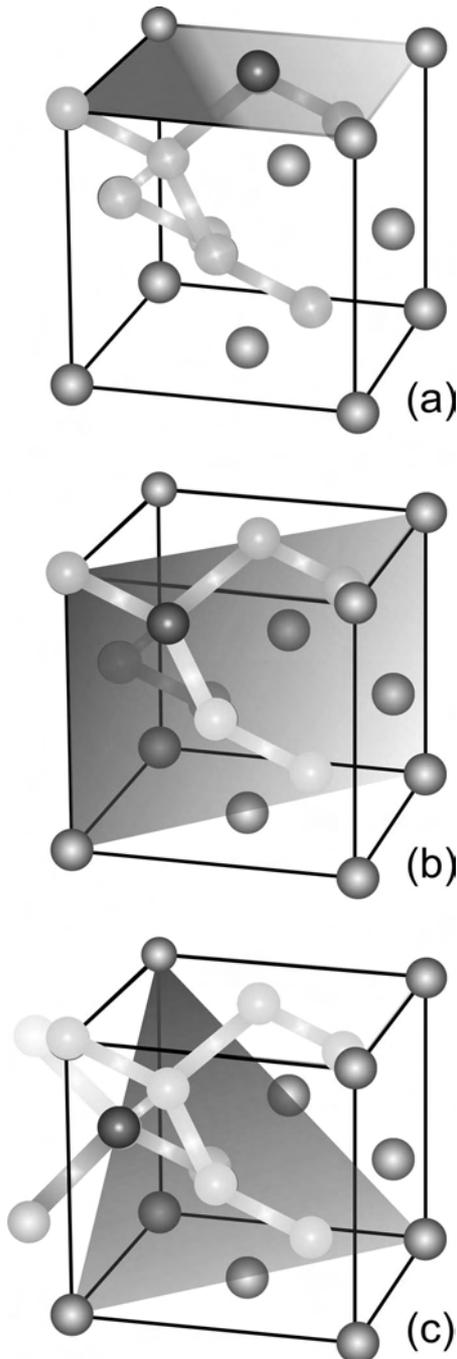

Fig. 4. The number of cells in the neighborhood an atom (black) is used to determine whether an atom belongs to a (100)-plane (a), (110)-plane (b) or (111)-plane (c).

ACKNOWLEDGMENT

The Deutsche Forschungsgemeinschaft (DFG) has financially supported this work within a collaborative Research Center (Sonderforschungsbereich 516) titled, 'Design and Fabrication of Active Microsystems'.

REFERENCES

[1] D. Straube, U. Triltsch, H.-J. Franke, S. Büttgenbach: *Modular software system for computer aided design of microsystems,* Microsystem Technologies, Vol. 12, 2006, April, S. 650-654
[2] O. Than and S. Büttgenbach: *Simulation of anisotropic chemical etching of crystalline silicon using a cellular automata model*, Sensors and Actuators A, Vol. 45, 1994, pp. 85-89
[3] U. Hansen, S. Büttgenbach: *T-CAD for the analysis and verification of* processing sequences, Microsystem Technologies 10, 2004, S. 193-198
[4] MEMSPro, SoftMEMS LLC, website: http://www.softmems.com, visited 13.02.2008
[5] CoventorWare, Coventor Inc., website: http://www.coventor.com, visited 13.02.2008
[6] S. D. Senturia, Microsystem Design, Kluwer Academic Publishers, 2001
[7] G. K. Fedder, "Top-Down Design of MEMS", Proc. of MSM 2000, 7 – 10, (2000)
[8] T. Mukherjee, G. K. Fedder, R. D. Blanton, "Hierarchical Design and Test of Integrated Microsystems", IEEE Design and Test Magazine, 18-27, (1999)
[9] U. Triltsch, S. Büttgenbach, D. Straube, H.-J. Franke: T-CAD Environment for Multi-Material MEMS Design, Proc. of Nanotech 2005, Vol. 3, Anaheim, USA, May 2005, S. 541-544
[10] U. Triltsch, A. Phataralaoha, S. Büttgenbach, D. Straube, H.J. Franke: Optimization of Lithographic Masks Using Genetic Algorithms, DTIP 2005, Montreux, CH, 2005 , June , S. 155-159
[11] M. A. Gosálvez, Atomistic Modelling of Anisotropic Etching of Crystaline Silicon, Dissertation of Laboratory of Physics, Helsinki University of Technology, 2008
[12] M. A. Gosálvez, VisualTAPAS: an example of density functional theory assisted understanding and simulation of anisotropic etching, J. Phys.: Condens. Matter 20, 2008
[13] J. Frühauf,. Shape & Functional Elements of Bulk-Silicon Microtechnique: Springer-Verlag Berlin, 2004

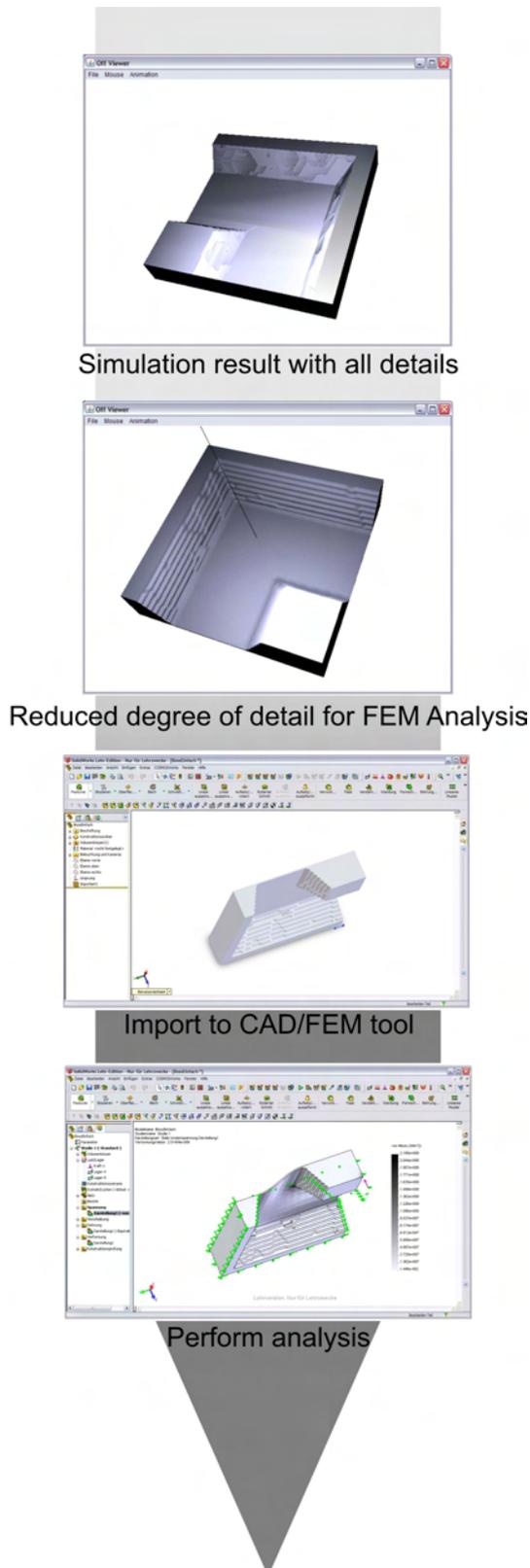

Fig. 5. The results can be exported any CAD or FEM tool using the ACIS-format.